\documentclass[iicol,sn-aps,Numbered]{sn-jnl}

\usepackage{graphicx}

\usepackage{multirow}%
\usepackage{amsmath,amssymb,amsfonts}%
\usepackage{amsthm}%
\usepackage{mathrsfs}%
\usepackage[title]{appendix}%
\usepackage{xcolor}%
\usepackage{textcomp}%
\usepackage{manyfoot}%
\usepackage{booktabs}%
\usepackage{algorithm}%
\usepackage{algorithmicx}%
\usepackage{algpseudocode}%
\usepackage{listings}%

\begin{document}
\title{Effects of ballistic transport on the thermal resistance and temperature profile in nanowires}

\author*[1]{R. Meyer}
\email{rmeyer@laurentian.ca}

\author[2]{Graham W. Gibson}

\author[1]{Alexander N. Robillard}

\affil[1]{Bharti School of Engineering and Computer Science, Laurentian University, 935 Ramsey Lake Road, Sudbury, ON\ \ P3E 2C6, Canada}
\affil[2]{Sudbury, Ontario, Canada}

\abstract{%
Effects of ballistic transport on the temperature profiles and thermal resistance in nanowires are studied. Computer simulations of  nanowires between a heat source and a heat sink have shown that in the middle of such wires the temperature gradient is reduced compared to Fourier's law with steep gradients close to the heat source and sink. In this work, results from molecular dynamics and phonon Monte Carlo simulations of the heat transport in nanowires are compared to a radiator model which predicts a reduced gradient with discrete jumps at the wire ends. The comparison shows that for wires longer than the typical mean free path of phonons the radiator model is able to account for ballistic transport effects. The steep gradients at the wire ends are then continuous manifestations of the discrete jumps in the model.
}

\date{\today}
\maketitle

\section{Introduction}
The development of nanotechnology and the miniaturization of devices has lead to a strong interest in the subject of  thermal transport at the nanoscale. As is the case with other physical properties, the laws describing thermal transport in macroscopic systems must be modified on the nanoscale. This is not only important for the design of nano-devices but can also be exploited to create nanomaterials with unusual thermal transport properties. Examples of technological areas that can benefit from an understanding of nanoscale thermal transport can be found in  \cite{Volz:2009,Maldovan:2014cb}.

Heat transport in macroscopic solid systems is well described by Fourier's law which is based on the idea of a diffusive transport mechanism where heat carriers are Brownian particles which undergo frequent collisions that lead to random direction changes. However, when the system dimensions are comparable to or shorter than the mean free path of the heat carriers between collisions, Fourier's law no longer provides an adequate description and microscopic transport mechanisms must be considered. 

In this work, we are using the simple case of a nanowire between a heat source and a heat sink to study deviations from Fourier's law caused by ballistic transport effects.  Results obtained from molecular dynamics simulations and phonon Monte Carlo simulations are compared to the prediction of Fourier's law and a simple phonon radiator model.

\begin{figure}
\centering
\includegraphics[width=7.5cm]{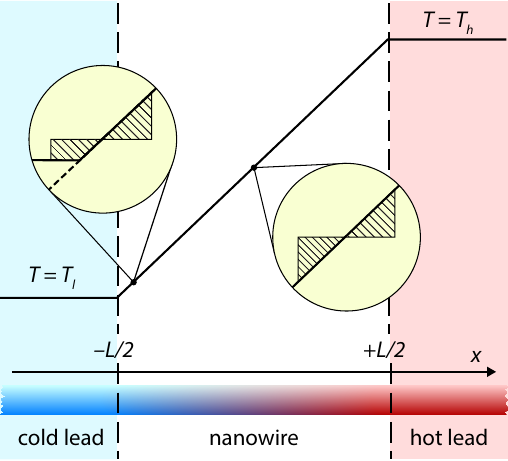}
\caption{Nanowire between two thermal leads. The upper part of the figure shows the temperature profile predicted by Fourier's law. The insets indicate the difference in the  local environments in the middle of the wire and close to the cold lead.}
\label{FigSketch}
\end{figure}

Figure~\ref{FigSketch} shows the type of systems we are considering in this work. A nanowire is connected to two thermal heat baths with temperatures $T_l$ and $T_h$. Fourier's law predicts that between the heat baths, the temperature follows a straight line that connects the temperatures of the two heat baths. 
 
Several studies of thermal transport through nanowires or thin films have shown deviations from the behaviour predicted by Fourier's law. Solutions of phonon radiator models \cite{Klitsner:1988, Majumdar:1993} predict a linear temperature profile with a reduced slope and discrete temperature jumps at the interface to the heat baths. A similar behaviour can be seen in results from a phonon Monte Carlo simulation (Fig.~4 in Ref.~\cite{Mazumder:2001in}). Molecular dynamics simulations \cite{Maiti:1997,Oligschleger:1999ck,Schelling:2002jl,Tang:2004gb,AbsDaCruz:2013fb,Li:2019} on the other hand show non-linear temperature profiles with steep slopes near the areas that are thermalized through contact with a heat bath (leads). Away from the leads the temperatures in the molecular dynamics simulations settle into a linear behaviour with a reduced slope compared to the Fourier case.
 
The deviations from Fourier's law observed in the previous studies can be understood qualitatively by looking at the insets in Fig.~\ref{FigSketch}.  The temperature profile in the immediate vicinity of any point in the middle of the wire (right inset) is symmetric so that the temperature increase to the right of the point is compensated by a similar decrease to its left. Further, the shape of the nearby profile remains the same everywhere in the middle of the wire. The situation changes, however, when we move the point close to one of the leads. As shown by the right inset, the temperature profile in the immediate vicinity is no longer symmetric and its shape depends on the distance to the lead. Since the temperature profile of the surroundings determines the distribution of incoming heat carriers, the behaviour of the temperature close to the leads has to account for the change in the environment.
  
The goal of this work is to show that the reduction of the temperature gradient inside the wires is a general phenomenon that is well described by a simple radiator model. To this end, we compare the analytical predictions of the model to results from molecular dynamics and phonon Monte Carlo simulations. This comparison shows that the radiator model provides a quantitative description of the reduction of the temperature gradient over a wide range of wire lengths.

The rest of the article is organized as follows. In Sec.~\ref{SecModel} we present the theory of the radiator model. (In the appendix we also present an extension of the model to a periodic geometry which facilitates the comparison to the molecular dynamics simulations). In Sec.~\ref{SecMD} and Sec~\ref{SecPMC} we compare the predictions of this model with results from molecular dynamics and phonon Monte Carlo simulations, respectively. The article finishes with a brief summary and conclusion in Sec.~\ref{SecSummary}.

\section{Model calculations}
\label{SecModel}
To understand the behavior of the temperature close to a thermalized lead, we consider a simple model for heat transport based on the idea introduced by P.~Debye in Ref.~\cite{Debye:1914}. In this model, some kind of heat carriers store and transport  heat energy by flowing through the material while undergoing random scattering events. The model makes no assumptions about the nature of the heat carriers and does not take into account different types of carriers such as phonons with different wavelengths or polarizations. A discrete version of a similar model was used by Klitsner et al.~\cite{Klitsner:1988}.

For a thin nanowire, we can treat the system as a quasi-one-dimensional problem so that there are only two possible directions for heat or energy flow along the wire axis. For simplicity, we refer to these two directions as left and right, independent of the orientation of the wire. For a thin slice of the wire at position $x$ we thus have to consider only the two energy fluxes $\Phi_\mathrm{l}(x)$ and $\Phi_\mathrm{r}(x)$ entering the slice from the left and right side, respectively.

Scattering reduces the energy fluxes as they pass through the slice. Assuming a Lambert absorption law the flux reductions are proportional to the magnitude of the fluxes and the width of the slice $\Delta x$. With an attenuation coefficient $\mu$, the total rate of energy going into scattering events, $S$,  in the slice at position $x$ is then
\begin{equation}\label{EqScatt}
	S(x) = \mu\,\Delta x\,\left[\Phi_\mathrm{l}(x)\,+\,\Phi_\mathrm{r}(x)\right].
\end{equation}
In the steady state there can be no net gain or loss of energy unless the slice is attached to a heat source or sink. Energy scattered from the incoming fluxes by scattering events is therefore re-emitted from the slice by adding the amount $\frac{1}{2} S(x)$ to the fluxes leaving the slice on either side.

If the attenuation coefficient $\mu$ is the same everywhere in the system, the Lambert absorption law leads to an exponential decrease of energy fluxes travelling through the system. Concretely, if $\frac{1}{2}S(x')$ is the flux emitted by a slice at $x'$ in one direction, only the amount
\begin{equation}\label{EqExp}
	\Phi(x, x') = S(x') e^{-\mu |x' - x|}
\end{equation}
reaches a slice located at $x$. Equation~\ref{EqExp} allows us to calculate the total fluxes $\Phi_\mathrm{l}(x)$ and $\Phi_\mathrm{r}(x)$ entering the slice at $x$. These two fluxes can be seen as the sum of contributions of an infinite series of slices to the left and right of the the slice, respectively. Taking the continuum limit of infinitesimally thin slices, and replacing the summation by integration, we obtain:
\begin{subequations}
\begin{eqnarray}
	\Phi_\mathrm{l}(x) &=& \frac{1}{2}\int_{-\infty}^{x} S(x') e^{-\mu |x-x'|} dx' \\
	\Phi_\mathrm{r}(x) &=& \frac{1}{2}\int_{x}^{\infty} S(x') e^{-\mu |x-x'|} dx'
\end{eqnarray}
\end{subequations} 
Combining these expressions with Eq.~\ref{EqScatt} we obtain the steady state equation 
\begin{equation}\label{EqPhi}
	S(x) = \frac{\mu}{2}\,\int_{-\infty}^{+\infty} S(x') e^{-\mu |x-x'|} dx' .
\end{equation}

To calculate temperature profiles, we need to connect the rate of energy being scattered by the heat carriers $S$ with the temperature $T$. Debye argued that the fluxes passing through a cross-section of the wire is proportional to the internal energy density of the material \cite{Debye:1914}. In the high temperature limit --- well  above the Debye temperature --- the internal energy is proportional to the temperature $T$. Since the rate of energy being scattered at $x$ is proportional to the fluxes at this point, we assume that $S(x) = \alpha T(x)$. Putting this into  Eq.~\ref{EqPhi} and removing  the proportionality constant $\alpha$ that appears on both sides, yields
\begin{equation}
\label{EqSteady}
	T(x) = \frac{\mu}{2}\,\int_{-\infty}^{+\infty} T(x') e^{-\mu |x-x'|} dx'  .
\end{equation}
This steady state equation determines the temperature at locations with no heat source or sink.
 
For a wire of length $L$, connected to infinitely long thermal leads with constant temperatures $T_l$ and $T_h$, it can be shown that Eq.~\ref{EqSteady} is solved by a linear profile
\begin{equation}
\label{EqAnsatz}
	T(x) = \left\{ \begin{array}{ll} 
		T_l & x \le -L/2 \\
		m\,x + \frac{(T_h + T_l)}{2} & -L/2 < x < L/2 \hspace{2em}\ \\
		T_h & x \ge L/2
	\end{array}\right. 
\end{equation}
with the temperature gradient
\begin{equation}
\label{EqSlope} 
	m = \frac{\mu}{2 + \mu L}\;(T_h - T_l) . 
\end{equation}
A proof of this solution is given in the appendix. 

\begin{figure}[t]
\centering
\includegraphics[width=7.5cm]{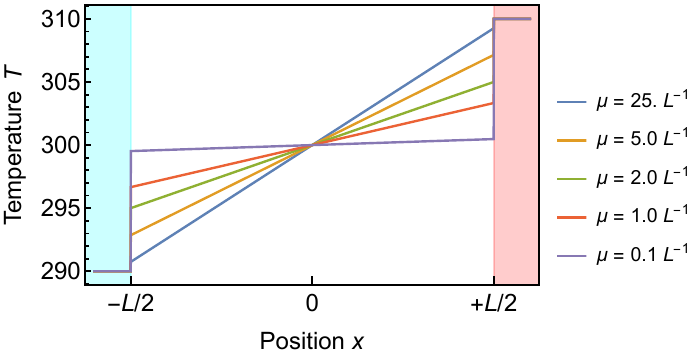}
\caption{Temperature profiles predicted by the radiator model for a wire of length $L$ and various attenuation constants $\mu$. The blue and red shaded areas indicate the position of the cold and hot thermal leads.}
\label{FigModelProfiles}
\end{figure}

Our simple model thus predicts a linear temperature profile similar to Fourier's law. However, the temperature gradient predicted by Eq.~\ref{EqSlope} is lower than in case of Fourier's law where the gradient is $\frac{T_h-T_l}{L}$. The lower gradient results in discontinuous temperature jumps at the interface between the leads and the wire. Figure~\ref{FigModelProfiles} shows the temperature profile for various values of the attenuation constant $\mu$. The size of the temperature jumps decreases as the length of the wire $L$ increases. In the limit of $L \rightarrow \infty$ the jumps vanish and Eqs.~\ref{EqAnsatz}, \ref{EqSlope} converge towards the prediction of Fourier's law.

The discontinuous temperature jumps predicted by the model are the necessary accommodations described in the introduction.  These jumps decrease the fluxes generated by the linear profile inside the wire so that they equal the fluxes entering from the constant temperature leads. Without these jumps, the change from the flat temperature profile in the thermalized areas to the linear profile in the middle part of the wire would lead to an inconsistency  between the temperature fluxes entering the system from the leads and the flux inside the wire. We continue the discussion of the discrete temperature jumps at the end of this section.

Further insight into the effect of the leads can be obtained by calculating of the net flux inside the wire as the difference between the fluxes going in the left and right direction: 
\begin{eqnarray}
	j(x) &=& \Phi_\mathrm{l}(x)\,-\,\Phi_\mathrm{r}(x) \nonumber \\
	&=& \alpha \int_{-\infty}^x T(x') e^{-\mu(x-x')} dx' \\
	& & -\;\alpha \int_x^{+\infty} T(x') e^{-\mu(x'-x)} dx' \nonumber 
\end{eqnarray}
Using Eqs.~\ref{EqAnsatz},\ref{EqSlope} we obtain for $-L/2 < x < +L/2$
\begin{equation}\label{EqFlux}
j(x) = \frac{\alpha}{2 + \mu L} (T_h - T_l) .
\end{equation}
The net flux is thus constant throughout the wire as it should be since there is no heat source or sink in this area.

From Eq.~\ref{EqFlux} we find the thermal conductance of the wire system
\begin{equation}\label{EqConductance}
	C = \frac{\alpha}{2 + \mu L} .
\end{equation}
The thermal resistance
\begin{equation}\label{EqResistance}
	R = C^{-1} = \frac{\mu}{\alpha} L\;+\;\frac{2}{\alpha} .
\end{equation}
is easier to interpret. Equation~\ref{EqResistance} shows that the resistance consists of two contributions. The first term is the thermal resistance of the wire material. It is proportional to the length $L$ of the wire with a bulk conductivity of $\kappa_{\infty} = \frac{\mu}{\alpha}$. The second term represents a constant resistance $R_c=\frac{1}{\alpha}$ that is added at the interface between the wire and the leads. Note that this is not the usual Kapitza resistance that occurs at interfaces between different materials. Our model assumes the same material for the lead and the wire. The only difference between the wire and the leads is that the leads are kept at constant temperatures. 

In summary, the simple model presented in this section predicts the occurrence of contact resistances at the interfaces between the  constant temperature leads and the linear temperature profile inside the wire. These contact resistances cause discrete jumps in the temperature at the interfaces. They reduce the net flux inside the wire so that it is consistent with the fluxes injected by the leads. This model is of course a simplified picture that neglects many details of the transport process such as different phonon wavelengths and polarizations. We expect that a more realistic treatment of the transport process would result in a smooth temperature profile without discrete jumps.

%
%
\section{Molecular dynamics simulations}
\label{SecMD}
\subsection{Details of the simulations}\label{SecMdDetails}
In this section we present results from a series of non-equilibrium molecular dynamics simulations of nanowires. The goal of these simulations was to understand to what extent the model calculations presented in the previous section predict the outcome of a more realistic treatment of the thermal transport on the nanoscale.

\begin{figure}
\centering
\includegraphics[width=7.5cm]{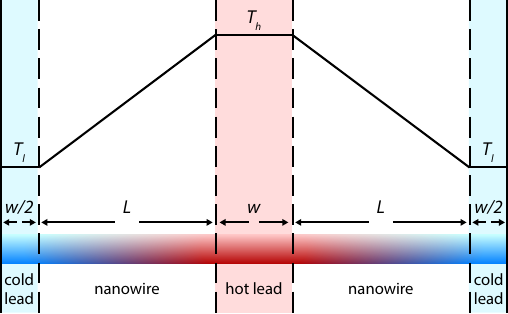}
\caption{Simulation cell used in the molecular dynamics simulations with periodic boundary conditions. The upper part of the figure shows the temperature profile predicted by Fourier's law.}
\label{FigMdSketch}
\end{figure}
Since molecular dynamics simulations are limited to finite system sizes, the simulations were carried out with finite simulation cells using periodic boundary conditions along the wire axis $x$. Along the $y$ and $z$ directions open boundary conditions were applied. As shown in Fig.~\ref{FigMdSketch}, the simulation cells contained two thermal leads of width $w$. Both types of leads (hot or cold) are connected to the other type by nanowire sections of length $L$ on their left and right side. Due to the mirror symmetry of the simulation cell, the temperature profile in the right half has to be a mirror image of the profile in the left half. In our analysis we therefore averaged the profiles of the two halves. As pointed out in the appendix, the model discussed in Sec. \ref{SecModel} can be also be solved for the periodic geometry used in the molecular dynamics simulations. The slope of the profile in the wire section for this geometry is given by Eq.~\ref{EqSlopeMd}.

The simulated wires were built by cutting a cylinder with a radius of 10 lattice constants from a perfect fcc lattice. The cylinder axis was aligned with the lattice's $[100]$ direction. For the interaction potential we chose the Lennard-Jones potential~\cite{LennardJones:1931}. To keep our results independent from any specific material, we set the energy and length parameters $\epsilon$ and $\sigma$ of the potential as well as the particle mass $m$ to unity. All parameters and results of the simulation are presented here using these dimensionless (reduced) units. All simulations were carried out with the simulation package LAMMPS~\cite{Plimpton:1995fc} using a time step $\Delta t = 0.005$.

Our first step was the determination of the equilibrium lattice constant of the wire systems along the wire axis. This step was necessary since the finite temperature as well as relaxations of the crystal lattice perpendicular to the wire axis lead to deviations of the lattice constant from its value at $T=0$. The lattice constant was determined from a simulation of a wire with a length of 200 lattice constants. During this simulation, the size of the system was relaxed with the help of the Parrinello-Rahman method~\cite{Parrinello:1980kx,Parrinello:1981}. The lattice constant $a_0=  1.55030$ determined from this simulation was then used in the following non-equilibrium simulations.

The main part of our molecular dynamics simulations was a series of simulations using wire systems of different length. The wire lengths in these simulations were: $L = 25\,a_0$, $50\,a_0$, $100\,a_0$, $200\,a_0$, $300\,a_0$, $400\,a_0$, $500\,a_0$, $600\,a_0$, $700\,a_0$, $800\,a_0$, $900\,a_0$, and $1000\,a_0$.  The width of the thermal leads was in all instances $w = 50 a_0$. The total number of atoms in these configurations ranges from 187,350 to 2,622,900 atoms. The thermal leads were realized in these simulations with the help of two Langevin thermostats set to temperatures $T_l = 0.045$ and $T_h = 0.055$. The damping constant of the thermostats was set to 1 time unit.

For the computation of temperature profiles in the simulated wires, we subdivided the configuration into slices with a thickness of $2.5\, a_0$ along the wire axis. During the simulations, the number of atoms in each slice and their total kinetic energy were measured each tenth simulation step and accumulated over periods of 1000 steps. At the end of each 1000 step period, the equipartition theorem was used to calculate a local temperature for each slice during the period. These data were written into a file which was used to monitor the progress of the simulation and to calculate profiles averaged over longer time spans. 

Prior to the measurement of the temperature profile, we simulated all systems until they had reached their steady state. In addition to a manual inspection of the temperature profiles, we used the time-evolution of the average slope of the wire section as the main criterion to decide when a system had reached the steady state. To this end, we averaged the temperature data over periods of 100,000 simulation steps and applied a linear fit to the temperature data in the wire section. The simulations were continued until the slope of the fit line no longer showed a systematic change and had settled into fluctuations around a constant value. 

%
%
\subsection{Temperature Profiles}\label{SecMdProfiles}
\begin{figure}
\centering
\includegraphics[width=7.5cm]{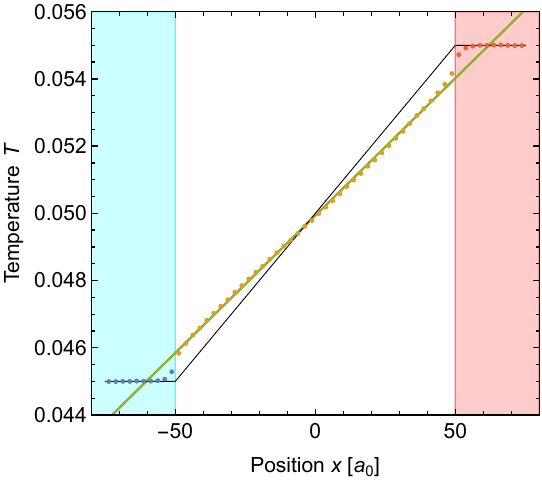}
\caption{Average thermal profile in a wire of length $L=100\,a_0$. The orange, blue and red dots indicate the temperatures in the wire, cold lead and hot lead sections of the system, respectively. The green line is the result of a linear fit of the data in the wire section. The thin black line shows the behaviour expected from Fourier's law. The blue and red shaded areas indicate the position of the cold and hot thermal leads.}
\label{FigMdProfile}
\end{figure}
After the model configurations had reached the steady state, the simulations were continued for another 1,000,000 steps. From the temperature data written to file during these simulations, the average temperature profiles of all configurations were determined. As an example, we show in Fig.~\ref{FigMdProfile} the profile calculated for the system with $L=100\,a_0$. The figure shows a very good agreement between the simulations and the predictions of the model presented in Sec.~\ref{SecModel}. There are sharp temperature jumps at the interface between the thermal leads and the wire. Apart from a small transition region close to these interfaces, the profile follows a straight line inside the wire and remains constant in the thermal lead sections. The temperature gradient in the simulation (green line) is clearly reduced compared to the prediction of Fourier's law (thin black line). A similar behaviour was observed for the systems with other wire lengths $L$. 

The presence of a transition area close to the interface between the thermal leads and the wire is to be expected for the same arguments we gave in the introduction. The thermal flux in the wire constantly adds or removes energy from the outer parts of the thermalized leads so that the environment there is different than in the centre of the leads. The thermostats can only heat or cool the leads at a certain rate, which depends on the thermostat settings. The temperature profile near the border of the leads is the result of a balance between the flux delivered by the thermostat and the flux going into or coming from the wire. A similar result might be expected in an experimental situation where the contact to a heat bath provides only a limited heating or cooling rate.

\begin{figure}
\centering
\includegraphics[width=7.5cm]{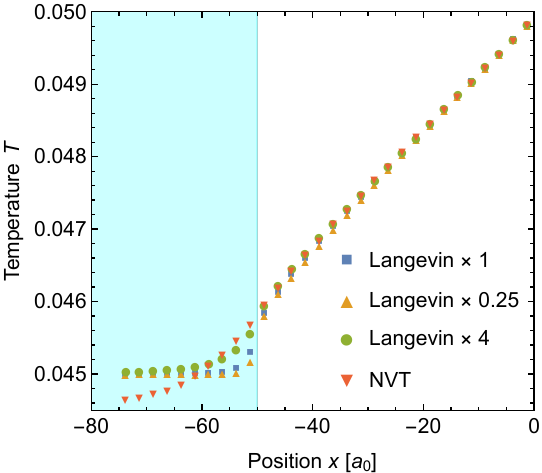}
\caption{Left half of the temperature profile in a wire of length $L=100\,a_0$ for different thermostat settings.}
\label{FigMdThermostats}
\end{figure}
In order to study the effect of the thermostat settings on the temperature profile, we have repeated the simulations of the wire of length $L=100\,a_0$ with the Langevin damping parameter scaled up and down by a factor of 4 as well as a Nose-Hoover NVT thermostat. The results of these simulations are given in Fig.~\ref{FigMdThermostats}. In order to show the differences between the setting clearly, we only show the transition at the lower temperature end of the system in this figure. The results at the higher temperature end are qualitatively the same.

The temperature profiles for different Langevin damping parameters in Fig.~\ref{FigMdThermostats} show that lower damping parameters result in a sharper temperature gradient at the interface. This agrees with the mechanism described in the last paragraph since lower damping parameters allow for a faster energy transfer between the thermostats and the systems.

Figure~\ref{FigMdThermostats} shows that the NVT thermostat behaves different than the Langevin thermostats. Apart from a very smooth entry into the thermostatted area, the temperature undershoots the set temperature of the thermostat (0.045) in the middle of the thermostatted region. This can be understood from the mechanics of this thermostat. The NVT thermostat does not drive the temperature  (more precisely its kinetic energy) of each atom towards the set value. Instead, the thermostat maintains the average temperature in the thermostatted region. The higher temperatures close to the interface in Fig.~\ref{FigMdThermostats} are therefore balanced by lower temperatures in the centre of the thermal lead.

While Fig.~\ref{FigMdThermostats} shows that the thermostat parameters affect the profile in the thermal leads, the figure also shows that there is little variation between the profiles inside the wire. Most importantly, the temperature gradient in the middle part of the wire is virtually the same in all four cases. This indicates that the deviations from Fourier's law are not caused by the dynamics of the thermostats.
 
\begin{figure*}[t]
\centering
\includegraphics[width=15.6cm]{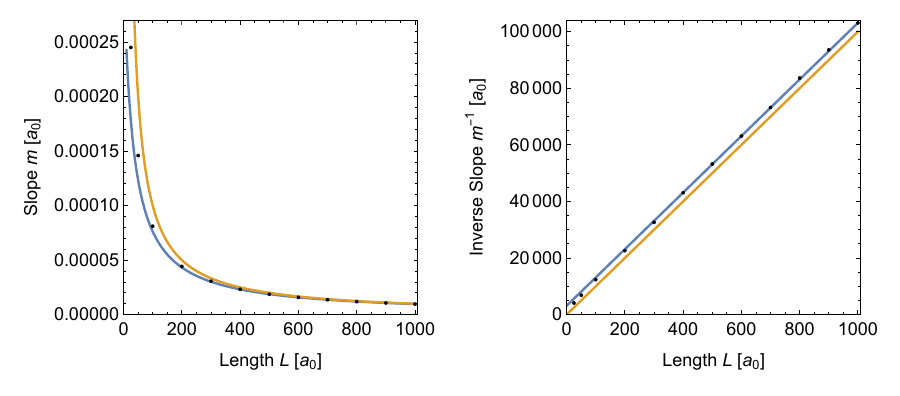}
\caption{Slope (left) and inverse slope (right) of the temperature profile in the wire section as a function of the wire length $L$. Black dots indicate the results from molecular dynamics simulations. Orange lines show the results expected from Fourier's law. The blue lines are the result of a fit of the inverse slope data with $L\ge 200\,a_0$.}
\label{FigMdSlope}
\end{figure*}
In order to make a quantitative comparison between our simulation results and the model predictions, we determined the slope of all profiles with the help of a linear regression of the temperature data in the wire sections. In order to minimize the effect of the transition region at the interfaces to the leads, the first and last data points were excluded from the fit.  In the left panel of Fig.~\ref{FigMdSlope} we present the resulting slopes or temperature gradients as a function of the wire length with the predictions of Fourier's law. It can be seen that the temperature gradients in the simulations are systematically below the prediction of Fourier's law (orange line) but converge towards the latter for longer wires. This is the expected behavior since Fourier's law is correct for macroscopic systems. Further details are revealed by the inverse slope data shown in the right panel of Fig.~\ref{FigMdSlope}. It can be seen that for longer wires the inverse slopes follow a straight line parallel to Fourier's law. This is in agreement with Eq.~\ref{EqSlopeMd} which predicts the inverse slope to be
\begin{equation}
\label{EqInvSlope}
	m^{-1} =  \frac{1}{T_h - T_l} \;L + \frac{2}{\mu\,(T_h - T_l)}\, \coth{\frac{\mu w}{2}}
\end{equation}
The blue lines shown in Fig.~\ref{FigMdSlope} were obtained by fitting the constant term in Eq.~\ref{EqInvSlope} to the simulation data for $L\ge 200\,a_0$. It corresponds to an attenuation constant $\mu =  0.0684$.

It can be seen from Fig.~\ref{FigMdSlope} that Eq.~\ref{EqSlopeMd} (represented by the blue fit line) provides a better description of the temperature profiles resulting from the simulations than Fourier's law. For shorter wires, however, Eq.~\ref{EqSlopeMd} underestimates the temperature gradients. In our opinion, this discrepancy is related to the fact that the temperature profiles in the simulations do not show a discrete temperature jump as predicted by the model in Sec.~\ref{SecModel}. Instead, they feature a thin but finite width transition region between the thermal leads and the wire (see for example Fig.~\ref{FigMdProfile}). Since the transition regions extend into the thermal leads, they reduce the need for adaptation in the wire section. Such a transfer of the adaptation is excluded in the model calculations which keep the lead temperatures constant as a boundary condition. The impact of the finite transition areas would be even larger for shorter wires since the temperature jump increases for shorter wires.

Another factor that might contribute to the differences between Eq.~\ref{EqSlopeMd} and the simulation data is our method to compute the slope values from the simulations. We carried out a number of tests where we changed the number of points excluded from the linear fit of the wire section. We also experimented with some non-linear fit functions that could account for the transition regions. None of these attempts lead to significant changes of our results. We are therefore convinced that the presence of the transition areas which is not included in the model calculations is the main reason for the increased temperature gradients in the simulations of shorter wires.

Overall, we believe that the results presented in this subsection confirm our basic premise that the temperature profile of the wire reacts to the different environment near the thermal leads. In accordance with our model calculations, the resulting accommodation at the wire ends result in a notable reduction of the temperature gradient at the center of the wire.

%
%
\subsection{Thermal resistances}\label{SecMdR}
\begin{figure}[b]
\centering
\includegraphics[width=7.5cm]{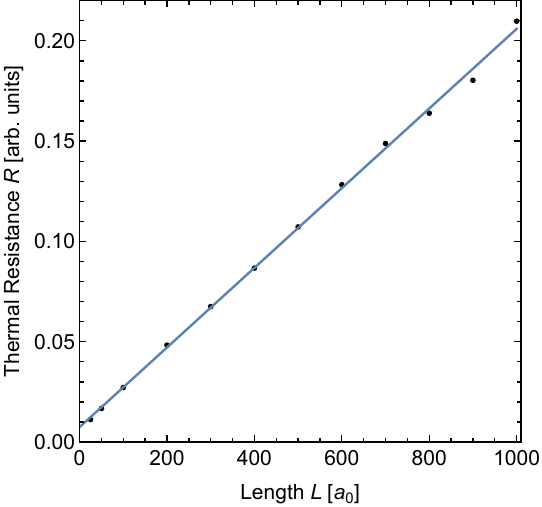}
\caption{Thermal resistance of the wire configurations as a function of the wire length. The blue line is a linear fit of the data.}
\label{FigMdFlow}
\end{figure}
In addition to the temperature profiles, we used the energy flux in the wires as a second quantity to compare the predictions of our model calculations to the molecular dynamics simulations. During the simulations, the cumulative amount of energy added to or subtracted from the system by the two thermostats that maintain the temperatures $T_h$ and $T_l$ in the lead sections were computed and periodically outputted (together with the temperature data). The difference of the two tallies corresponds to the total amount of energy that has been transported through the wire sections during the simulation. Over the course of the simulation, this value increases on average linearly with time. By fitting a straight line to the data and taking its slope, we were thus able to determine the average net energy flux running through the wires and from these the thermal resistance $R$ for each wire configuration.

In Fig.~\ref{FigMdFlow} we present the thermal resistances $R$ determined from the molecular dynamics simulations as a function of the wire length $L$. This figure agrees qualitatively very well with Eq.~\ref{EqResistance} which predicts that the thermal resistance is a linear function of the wire length with a non-zero offset. A complete quantitative comparison between the predictions of  Eq.~\ref{EqResistance} and the simulation data is not possible since the model does not yield information about $\alpha$. However, according to  Eq.~\ref{EqResistance} the attenuation constant $\mu$ equals twice the slope of the fit line divided by its offset. Applying this method to the values of the fit line in Fig.~\ref{FigMdFlow} gives a value of $0.0532 a_0^{-1}$ which is about 22\,\% smaller than the value $\mu = 0.0684 a_0^{-1}$ obtained from the temperature profiles.

In our opinion, the difference between the $\mu$ values is a result of the fact that the model used to derive Eq.~\ref{EqResistance} does not account for the details of the transport mechanism. We believe, however, that  Fig.~\ref{FigMdFlow} provides support for the general prediction of the model which is that the interfaces between the thermal leads and the wire give rise to a constant interface resistance which induces steep temperature changes in the vicinity of the interface. It should be noted that the results in this subsection are independent from the results obtained in the preceding subsection since the cumulative energy tallies of the thermostats are not affected by the local temperature distribution inside the wire.

\section{Phonon Monte Carlo simulations}
\label{SecPMC}
\subsection{Details of the simulations}\label{SecPMCDetails}
The results presented in the preceding section show a good agreement between the molecular dynamics simulations and the predictions of the model from Sec.~\ref{SecModel}. In this section we present results from phonon Monte Carlo simulations as an independent test of the validity of the model predictions. By using a completely independent simulation method we are able to remove any remaining doubt whether the results in the preceding section are artefacts of the artificial dynamics of the thermostats in the molecular dynamics simulations.

The phonon Monte Carlo simulation method is a relatively new method for the simulation of heat transport in non-metallic materials. The method was first introduced by Mazumder and Majumdar \cite{Mazumder:2001in} based on preliminary work by Peterson \cite{Peterson:1994cu}. Further improvements were published by Lacroix, Joulain and Lemmonier \cite{Lacroix:2005gm} as well as P{\'eraud, Hadjiconstantinou and Nicolas \cite{Peraud:2011ks,Peraud:2012bj}. In a phonon Monte Carlo simulation, packages of phonons  are propagated through the simulated system. During the simulation, the phonons in the packages undergo stochastic collisions leading to changes of the direction and momentum (frequency) of the phonons. Details of the method can be found in Refs.~\cite{Mazumder:2001in,Lacroix:2005gm,Peraud:2011ks,Peraud:2012bj,Raleva:2017}. Details of the implementation of the phonon Monte Carlo method used in this work can be found in \cite{Gibson:2022}.

Phonon Monte Carlo simulations require a scattering model that describes the lifetime of different phonon modes. In this work we have used two different such models: the more recent parametrization used by Jean et al.~\cite{Jean:2014gw} (\textit{Si-1}) and the older parametrization used by Lacroix et al.~\cite{Lacroix:2005gm} which is based on data by Holland (\textit{Si-2}). The usage of the two models allows us to study the effect of the phonon mean free path since the latter model is known to predict shorter phonon life times and therefore shorter mean free paths than the former. Both models only account for the phonon scattering rates due to phonon collisions. It would be possible to add other scattering mechanisms such as impurity scattering and/or partially diffusive surface scattering to the simulations. These effects, in particular diffusive surface scattering, are certainly important for a quantitative description of thermal transport in nanowires. However, for the purpose of this work which is to study the effects of the thermal leads, the difference between \textit{Si-1} and \textit{Si-2} is sufficient. A more detailed study of the effects of different scattering mechanisms is beyond the scope of this work.

Using the two scattering models, we performed simulations of straight wires with lengths $L$ in the range from $1\,\mu m$ to $50\,\mu m$ in the case of \textit{Si-2} and $0.1\,\mu m$ to $10\,\mu m$ in the case of \textit{Si-2}. The wire width was in all cases 20\,nm. It should be noted that the simulation code only allows the specification of a two-dimensional geometry. The z-dimension is treated as an infinite open system. Apart from a scaling factor, the results obtained in this manner are, however, equivalent to a system that is limited along the z direction by perfectly specular boundaries. Our results therefore correspond to simulations of a square wire with 20\,nm side length. For all wire lengths, we simulated 500,000,000 phonon packages. with a maximum lifetime of $400\,\mathrm{ns}$ per $\mathrm{\mu m}$ of system length, $L$.

\subsection{Temperature Profiles}\label{SecPmcProfiles}
\begin{figure*}
\centering
\includegraphics[width=15.6cm]{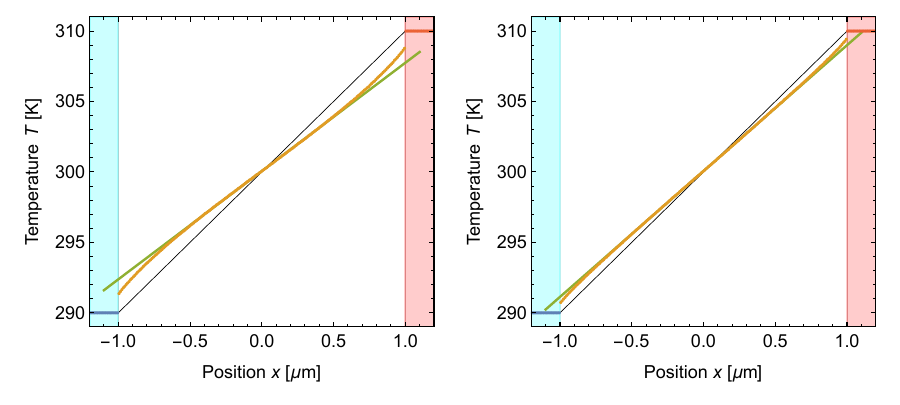}
\caption{Average thermal profile in a Si wire of length $L=2\,\mathrm{\mu m}$ from phonon Monte Carlo simulations  using the scattering models \textit{Si-1} (left) and \textit{Si-2} (right). Temperatures in the wire, the hot lead and the cold lead are shown in orange, red and blue, respectively. The green lines are result of linear fits of the data in the center half of the wires. The thin black lines shows the behaviour expected from Fourier's law. The blue and red shaded areas indicate the position of the cold and hot thermal leads.}
\label{FigPmcProfile}
\end{figure*}
As an example, Fig.~\ref{FigPmcProfile} shows the temperature profiles of a Si wire with a  length $L$ of $2\,\mathrm{\mu m}$ obtained from phonon Monte Carlo simulations for the two scattering models. It should be noted that the phonon Monte Carlo simulations do not simulate the thermalized leads but consider only phonons emitted from these areas into the wire (or absorbed by them). The lead areas in Fig.~\ref{FigPmcProfile} are therefore not simulation results but are shown in order to visualize the change of the profile, most notably the jumps at the interfaces.

The temperature profiles in Fig.~\ref{FigPmcProfile} are notably curved and follow a straight line only in the middle part of the wire. Although this effect is also present in Fig.~\ref{FigMdProfile}, it is clearly much more pronounced here. However, many molecular dynamics studies have also found curved profiles near thermostatted regions (e.g. Refs.~\cite{Maiti:1997,Oligschleger:1999ck,Schelling:2002jl,Tang:2004gb,AbsDaCruz:2013fb,Li:2019}). We believe that the extent of the curvature is determined by the mean free path of the heat carriers. As we pointed out in the introduction, some accommodation is expected near a thermostatted region and the longer the mean free path of the heat carriers, the longer such accommodations should reach into the wire. It is known that phonons with mean free paths above one $\mathrm{\mu}$m contribute significantly to the heat transport in Si \cite{Esfarjani:2011,Regner:2013}. This is consistent with the extent of the curvature shown by both models in Fig.~\ref{FigPmcProfile}.

Due to the curvature of the temperature profiles, it is not possible to fit the complete profiles obtained from phonon Monte Carlo simulations to a straight line} in order to derive the slope in the center. The green line in Fig.~\ref{FigPmcProfile} was instead obtained by fitting a portion with length $L/2 = 1\,\mathrm{\mu m}$ at the center of the wire. The same procedure was used to determine the profile slopes for other wire lengths.

\begin{figure*}
\centering
\includegraphics[width=15.6cm]{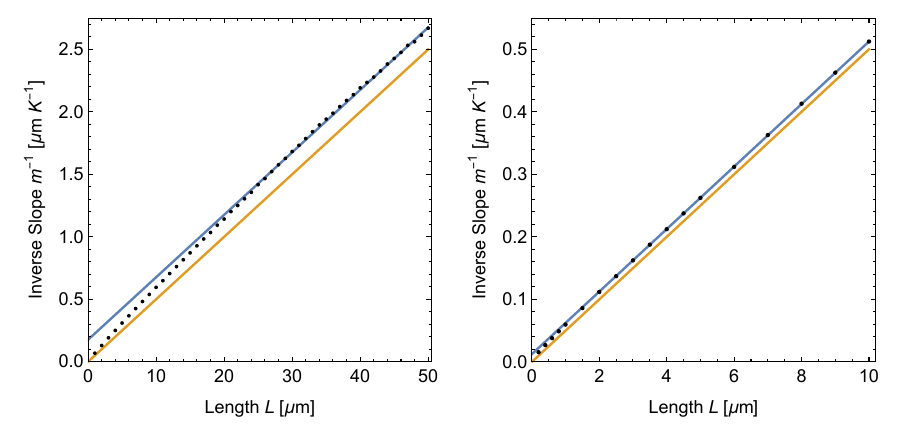}
\caption{Inverse slope of the temperature profile in Si wires as a function of the wire length $L$ obtained from phonon Monte Carlo simulations using the scattering models \textit{Si-1} (left) and \textit{Si-2} (right). Black dots indicate the simulation results. The orange lines shows the behaviour expected from Fourier's law. The blue line is the result of a fit of the inverse slope data for wires with $L\ge20\,\mathrm{\mu m}$ (left) and $L\ge2\,\mathrm{\mu m}$ (right).}
\label{FigPmcSlope}
\end{figure*}
In Fig.~\ref{FigPmcSlope} we present the inverse slopes obtained from phonon Monte Carlo simulations with the two scattering models.  In the case of \textit{Si-1} the inverse slopes follow a line parallel to the prediction by Fourier's law for wires longer than about $20 \mu m$. For shorter wires, the slopes begin to diverge from this line. Similarly, the inverse slopes obtained with model \textit{Si-2} follow a straight line for wires longer than about $2 \mu m$ and diverge from this line for shorter wires. Qualitatively, this is the same behavior shown by the molecular dynamics simulations (cf. Fig.~\ref{FigMdSlope}). As discussed above, we believe that the divergence from the straight line for shorter wires is a result of the fact that the profiles are not simply straight lines with jumps at the interfaces. As shown by the curvatures in Fig.~\ref{FigPmcSlope} the temperature profiles reach a constant gradient only at a certain distance into the wire. For shorter wires the accommodations from both ends overlap resulting in a higher slope.

\subsection{Thermal resistances}\label{SecPmcR}
\begin{figure*}
\centering
\includegraphics[width=15.6cm]{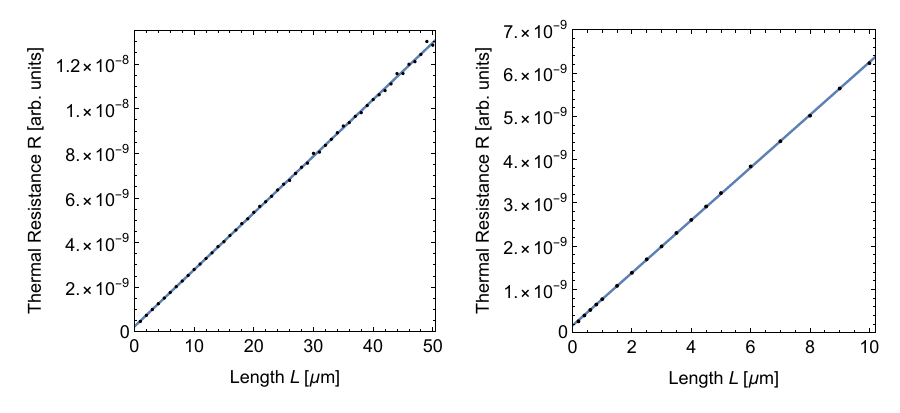}
\caption{Thermal resistance of Si nanowires as a function of the wire length obtained from phonon Monte Carlo simulations  using the scattering models \textit{Si-1} (left) and \textit{Si-2} (right).} The blue line is a linear fit of the data.
\label{FigPmcFlux}
\end{figure*}

The phonon Monte Carlo method also allows the calculation of the heat flux in the wires which we use again as an independent test quantity. Fig.~\ref{FigPmcFlux} shows the thermal resistances obtained from the heat fluxes in our simulations of Si nanowires. The figure shows that for both scattering models the resistances follow a straight line with a non-zero offset. Although this offset appears small in the figures, it adds a significant amount of resistance to shorter wires. This behavior is similar to the results obtained from molecular dynamics simulation (cf. Fig.~\ref{FigMdFlow}).  It is interesting to note that both simulation methods do not show noticeable deviations from a straight line for shorter wires. This means that the resistance added by the interface with the thermostatted regions remains unaffected by the mechanism leading to changes in the thermal gradients in shorter wires.

The values of the attenuation coefficient  $\mu$ obtained from the offset of the fit lines in Fig.~\ref{FigPmcFlux} are $1.05\,\mathrm{\mu m^{-1}}$ for model \textit {Si-1} and $3.69\,\mathrm{\mu m^{-1}}$ for \textit {Si-2}. These values are substantially smaller than the values $11.4\,\mathrm{\mu m^{-1}}$ and $161\,\mathrm{\mu m^{-1}}$ obtained from the fit lines in Fig.~\ref{FigPmcSlope} for the two models. This leads to the conclusion that the radiator model provides a qualitatively correct picture but is not able to predict the value of the constant resistance offset based on a single attenuation constant alone. A quantitative prediction of the offset probably needs to take into account the details of the mechanisms leading to the curved temperature profiles at the ends of the wires.

\section{Summary and conclusions}
\label{SecSummary}
In this work we have carried out non-equilibrium molecular dynamics simulations and phonon Monte Carlo simulations of thermal transport through nanowires. The results from these simulations are used to study the impact of ballistic transport effects on the temperature profiles and the thermal resistance of nanowires and similar quasi one-dimensional nanostructures. We also present a simple phonon radiator model. This model provides insight into the behavior observed in the simulations and an understanding of the underlying physical mechanisms.

In accordance with previous studies \cite{Maiti:1997,Mazumder:2001in,Oligschleger:1999ck,Schelling:2002jl,Tang:2004gb,AbsDaCruz:2013fb,Li:2019}, our computer simulations show temperature gradients in the centre of the system that are reduced compared to the prediction of Fourier's law. In order to achieve the full temperature difference over the system length the temperature profiles are curved with steeper gradients at the end of the system close to the thermalized leads. The degree of curvature of the profiles varies between the simulations.  We attribute the variation of the curvature in the simulations to differences in the phonon mean free path. It seems reasonable that in a system with a longer mean free path the effects of the thermal leads reach deeper into the system.

The radiator model predicts exactly linear temperature profiles with discrete jumps at the end of the system. While this is not exactly the behavior shown by the simulations, we believe that the steep gradients and curved profiles in the simulations are a continuous manifestation of the discrete jumps. Both simulation methods account for a range of phonon modes with different mean free paths whereas the radiator model only accounts for a single carrier particle with a uniform attenuation constant. Although the model does not exactly predict the behavior in the simulations, it provides a physical explanation for the reduction of the temperature gradients. Without this reduction the thermal flow in the wire does not match the flows generated by the thermal leads.

The inverse temperature gradients (profile slopes) in the centre of the wires obtained from our simulations follow a straight line for longer wires. This behavior is predicted by the phonon radiator model. For shorter wires the inverse gradients diverge from the straight line. A likely explanation for this are the curved tails of the temperature profile which reach a finite distance into the wires. For shorter wires the tails from both ends will begin to overlap resulting in an increased temperature gradient.

In addition to the temperature gradients, we determined the thermal resistance of the systems from our simulations. As predicted by the radiator model, the results follow a straight line with a non-zero offset for the full range of system lengths. The offset represents a constant thermal resistance caused by the interface to the thermal leads. While the variation of the thermal resistance is qualitatively predicted by the radiator model, it appears that the model is not able to predict the size of the resistance offset. Similar to the temperature gradients, a quantitative prediction would probably require a more sophisticated model.

The addition of a constant resistance to the wires by the thermal leads is similar to the well-known Kapitza effect which causes a constant thermal resistance at the interface between materials. One might ask whether the artificial dynamics of the thermostats in the molecular dynamics simulations modify the vibrational properties in the thermostatted regions sufficiently to cause a Kapitza resistance. However, if this was true, why do the phonon Monte Carlo simulations (as well as the radiator model)  show the same effect? In the phonon Monte Carlo simulations there is no difference between the materials in the thermal leads and the wire. Phonons enter the wire from the thermal leads with the expected frequency distribution and without any scattering so that there is no reason for a Kapitza resistance. Furthermore, our tests with different thermostat settings show that the temperature profile in the wire is unaffected by the details of the thermostat dynamics. This also speaks against the Kapitza mechanism as a cause for the additional resistance.

In summary, our simulations show the impact of ballistic transport on the heat conduction in nanowires. With some limitations, the radiator model is able to account for these deviations. For wires much longer than the typical mean free path of phonons, the model predicts merely a constant offset of the resistance which becomes negligible for wires of macroscopic length. For wires that are closer to the Casimir limit \cite{Casimir:1938}, i.e. wires that are much shorter than the typical mean free path of phonons, a more refined model that captures details of the scattering mechanism is required.

In our opinion, the ballistic transport effects studied in this work might be relevant for two reasons. The first reason is the additional term that appears in the thermal resistance. The second reason is the non-constant temperature gradient which might be important in applications that are sensitive to the local value of the temperature gradient. Both the steep gradient close to the thermal leads as well as the reduced gradient elsewhere might have to be considered in such applications.

\backmatter

%
%
\bmhead{Acknowledgments}
This work was supported financially by the Natural Sciences and Engineering Research Council of Canada (NSERC), grant number RGPIN/6563-2018, Laurentian University and the Ontario Graduate Scholarship program. The authors gratefully acknowledge the generous allocation of computational resources by the Digital Research Alliance of Canada.   

\bmhead{Availability of Data}
The raw data sets generated from molecular dynamics or phonon Monte Carlo simulation in the course of this work are available from the corresponding author upon reasonable request.

\bmhead{Author Contribution Statement}
Ralf Meyer wrote the initial draft of the manuscript and created the artwork and calculated the solutions of the radiator model. He designed the study and formulated the radiator model as used in this work in conjunction with Alexander Robillard who performed the molecular dynamics simulations. Graham developed and ran the computer program for the Phonon Monte-Carlo simulations. All authors provided critical comments on the manuscript. The final version of the manuscript was read and approved by all authors.

\begin{appendices}
%
%
\section{Solution for the infinite thermal leads}
\label{AppInf}
In this appendix we show that in the case of infinite thermal leads, Eqs.~\ref{EqAnsatz} and \ref{EqSlope} solve the steady state equation~\ref{EqSteady} of our model. We begin by noting that it is sufficient to treat the symmetric system where $T_h = -T_l = \Delta T$. Since the equality
\begin{displaymath}
T_0 = \int_{-\infty}^{+\infty} T_0 e^{-\mu |x-x'|} dx' ,
\end{displaymath}
holds for any constant temperature value $T_0$, the solution for general values $T_h$ and $T_l$ can be found by adding the offset $T_0 = \frac{T_h + T_l}{2}$ to the solution of the symmetric case with $\Delta T= \frac{T_h - T_l}{2}$.

To facilitate the calculations, we split the integral on the right hand side of Eq.~\ref{EqSteady} into four terms:
\begin{equation}\label{EqSteady4}
T(x) = e^{-\mu \left(\frac{L}{2}+x\right)} l_- + w_- + w_+ + e^{-\mu \left(\frac{L}{2}-x\right)} l_+ .
\end{equation}
Here 
\begin{subequations}
\begin{eqnarray}
	w_- &=& \frac{\mu}{2}\int_{-L/2}^{x} T(x') e^{-\mu (x-x')} dx' \\
	w_+ &=& \frac{\mu}{2}\int_{x}^{L/2} T(x') e^{-\mu (x'-x)} dx'
\end{eqnarray}
\end{subequations} 
are the contributions to the integral from inside the wire portion of the system whereas the other two terms represent the contributions from the  thermal leads. The latter terms consist of the integrals
\begin{eqnarray*}
	l_- &=& \frac{\mu}{2}\int_{-\infty}^{-L/2} T(x') e^{\mu \left(x'+\frac{L}{2}\right)} dx' \\
	l_+ &=& \frac{\mu}{2}\int_{L/2}^{\infty} T(x') e^{-\mu \left(x'-\frac{L}{2}\right)} dx'
\end{eqnarray*}
which represent the heat flux entering the wire from the left and right side, respectively, and an exponential factor accounting for the attenuation of these fluxes inside the wire. 

It is easy to see that $T_{h,l} = \pm\Delta T$  leads to $l_+ = -l_-$. This allows the simplification
\begin{equation}\label{EqSumL}
e^{-\mu \left(\frac{L}{2}+x\right)} l_- + e^{-\mu \left(\frac{L}{2}-x\right)} l_+ = 2\,\sinh{\left(\mu x\right)}\, e^{-\mu\frac{L}{2}}\, l_+.
\end{equation}

With the linear Ansatz $T(x) = m x$ inside the wire, the integrals $w_-$, $w_+$ can be evaluated, yielding
\begin{eqnarray*}
	w_- &=& \frac{m x}{2} - \frac{m}{2 \mu} + \frac{m}{4 \mu} e^{-\mu \left(\frac{L}{2}+x\right)} (2 + \mu L)\\
	w_+ &=& \frac{m x}{2} + \frac{m}{2 \mu} - \frac{m}{4 \mu}  e^{-\mu \left(\frac{L}{2}-x\right)} (2 + \mu L)
\end{eqnarray*}
and
\begin{equation}\label{EqSumW}
	w_- + w_+ = m x - \frac{m}{2 \mu} \sinh{\left(\mu x\right)}\, e^{-\mu\frac{L}{2}}\,(2 + \mu L)
\end{equation}

Putting the Ansatz $T(x) = m x$ and Eqs.~\ref{EqSumL}, \ref{EqSumW} into the steady state equation Eq.~\ref{EqSteady4} gives
\begin{equation}\label{EqSteadySolved}
m x = m x + \sinh{\left(\mu x\right)}\, e^{-\mu\frac{L}{2}}\,\left(2 l_+ - \frac{m}{2 \mu} (2 + \mu L)\right).
\end{equation}
This shows the the linear Ansatz fulfills the steady state equation inside the wire if and only if the second term on the right side vanishes for all $x$. This means that the slope $m$ must be a root of the equation
\begin{equation}\label{EqSlopeCondition}
2 l_+ - \frac{m}{2 \mu} (2 + \mu L) = 0.
\end{equation}

Using the value 
\begin{displaymath}
l_+ = \frac{\mu}{2}\int_{-\infty}^{-L/2} \Delta T e^{\mu \left(x'+\frac{L}{2}\right)} dx' = \frac{\Delta T}{2}
\end{displaymath}
it is easy to show that Eq.~\ref{EqSlope} solves Eq.~\ref{EqSlopeCondition} for the case of infinite thermal leads.

%
%
\section{Solution for the periodic case}
As discussed in Sec.~\ref{SecMD} and shown in Fig.~\ref{FigMdSketch}, the molecular dynamics simulations use a periodic cell containing two wire segments of length $L$ and two thermal leads of width $w$. For the same reasons given in Appendix~\ref{AppInf}, it is sufficient to treat the symmetric case $T_h = -T_l = \Delta T$.

For our calculations, we assume that the point $x=0$ aligns with the midpoint of a wire segment with leads at temperature $T_l$ and $T_h$ attached to it on its left and right side, respectively. On the interval $-\frac{L}{2}  < x < +\frac{L}{2}$ we then make the same Ansatz $T(x) = m x$ as in the case of infinite leads. The symmetry of the system then dictates that the temperature in the wire sections with a lead at $T_h$ on its left side is also a straight line with the slope $-m$.

 With the point $x$ being the mid point of one of the wires and $T_h = -T_l = \Delta T$, our Ansatz is an odd function of $x$, i.e. $T(x) = -T(-x)$. From this it follows that the lead integrals $l_-$, $l_+$ fulfill again the condition $l_+ = -l_-$. Further, since we are making the same Ansatz for the temperature on the interval $\frac{L}{2}  < x < +\frac{L}{2}$, the calculations used in Appendix~\ref{AppInf}  to derive Eqs.~\ref{EqSteadySolved} and \ref{EqSlopeCondition} also apply to the periodic geometry. The difference between the two systems is the value of $l_+$. By making use of the fact that the temperature is an antiperiodic function of $x$ with $T(x) = -T(x+L+w)$ the value of $L_+$ for the periodic geometry can be found to be:
\begin{equation}
l_+ = \frac{\Delta T \left(1-e^{-\mu  w}\right) + \frac{m \left((2 + \mu  L) e^{-\mu (L+w)}-(2-\mu  L) e^{-\mu  w}\right)}{2 \mu } }{2 \left(e^{-\mu (L+w)}+1\right)}
\end{equation}
Putting this result into equation \ref{EqSlopeCondition} and solving for $m$ then gives the slope of the temperature in the wire sections for the periodic geometry:
\begin{equation}
\label{EqSlopeMd}
	m = \frac{\mu}{2 \coth{\frac{\mu w}{2}} + \mu L}\;(T_h - T_l) . 
\end{equation}

Since $\coth\frac{\mu w}{2} > 1$ (for positive $\mu$ and $w$), Eq.~\ref{EqSlopeMd} shows that compared to the $\Delta T / L$ behavior of Fourier's law, the slopes in the periodic geometry are even more reduced than in the case of infinite thermal leads. Physically, this can be understood, since close to the end of a wire, the system will not only be influenced by the constant temperature in the thermal lead but also by the temperature profile in the wire section on the other end of the lead. This makes the hot (cold) lead appear colder (warmer) to the wire sections, explaining the reduced slopes. The behavior of the $\coth{}$ function results in a vanishing slope in the case of $w \rightarrow 0$ whereas the result of the infinite lead case (Eq.~\ref{EqSlope}) is recovered for $w \rightarrow \infty$. 

%
%
\section{Calculation of net fluxes}
\label{AppFlux}
The results obtained in the preceding sections can be used to calculate the net energy flux inside the wires. The net flux at any point $x$ in the wires is given by the difference between the fluxes arriving at $x$ from the left and right side. The energy reemitted from $x$ does not contribute to the net flux since equal amounts of energy are emitted in both directions. Using the integrals $w_-$, $w_+$, $l_-$, $l_+$ this yields
\begin{equation}\label{EqJ}
j(x) = \alpha \left(e^{-\mu \left(\frac{L}{2}+x\right)} l_- + w_- - w_+ - e^{-\mu \left(\frac{L}{2}-x\right)} l_+\right) .
\end{equation}
We have reinstated here the proportionality factor $\alpha$ between temperatures and energy fluxes which had been canceled from the steady state equation when going from Eq.~\ref{EqPhi} to Eq.~\ref{EqSteady}. 

Using the results obtained above, Eq.~\ref{EqJ} can be simplified to 
\begin{equation}
j(x) = - \frac{\alpha\,m}{\mu} - \alpha \cosh{(\mu x)} \, e^{-\mu\frac{L}{2}}\,\left(2 l_+ - \frac{m}{2 \mu} (2 + \mu L)\right).
\end{equation}
Since $m$ solves Eq.~\ref{EqSlopeCondition}, the second term on the right hand side vanishes so that 
\begin{equation}
j(x) = - \frac{\alpha\,m}{\mu}\,.
\end{equation}
This becomes Eq.~\ref{EqFlux} in the case of infinite leads and 
\begin{equation}
j(x) = \frac{\alpha}{2 \coth{\frac{\mu w}{2}} + \mu L}\;(T_h - T_l)
\end{equation} 
for the periodic systems.

\end{appendices}

%
%

\begin{thebibliography}{10}
\providecommand{\url}[1]{{#1}}
\providecommand{\urlprefix}{URL }
\providecommand{\doi}[1]{\url{https://doi.org/#1}}
\bibcommenthead

\bibitem{Volz:2009}
S.~Volz,  (Springer, Heidelberg, 2009), \emph{Topics in Applied Physics}, vol.
  118, chap.~1

\bibitem{Maldovan:2014cb}
M.~Maldovan, {Sound and heat revolutions in phononics}.
\newblock Nature \textbf{503}(7475), 209--217 (2013)

\bibitem{Klitsner:1988}
T.~Klitsner, J.E. VanCleve, H.E. Fischer, R.O. Pohl, Phonon radiative heat
  transfer and surface scattering.
\newblock Phys. Rev. B \textbf{38}, 7576 (1988)

\bibitem{Majumdar:1993}
A.~Majumdar, Microscopic heat conduction in dielectric thin films.
\newblock J. Heat Transfer \textbf{115}, 7 (1993)

\bibitem{Mazumder:2001in}
S.~Mazumder, A.~Majumdar, {Monte Carlo Study of Phonon Transport in Solid Thin
  Films Including Dispersion and Polarization}.
\newblock J. Heat Transfer \textbf{123}, 749--759 (2001)

\bibitem{Maiti:1997}
A.~Maiti, G.D. Mahan, S.T. Pantelides, Dynamical simulations of nonequilibrium
  processes -- heat flow and the {K}apitza resistance across grain boundaries.
\newblock Solid State Commun. \textbf{102}, 517--521 (1997)

\bibitem{Oligschleger:1999ck}
C.~Oligschleger, J.C. Sch{\"o}n, {Simulation of thermal conductivity and heat
  transport in solids}.
\newblock Physical Review B \textbf{59}(6), 4125--9 (1999)

\bibitem{Schelling:2002jl}
P.K. Schelling, S.R. Phillpot, P.~Keblinski, {Comparison of atomic-level
  simulation methods for computing thermal conductivity}.
\newblock Physical Review B \textbf{65}, 144306 (2002)

\bibitem{Tang:2004gb}
Q.~Tang, {A molecular dynamics simulation: the effect of finite size on the
  thermal conductivity in a single crystal silicon}.
\newblock Molecular Physics \textbf{102}(18), 1959--1964 (2004)

\bibitem{AbsDaCruz:2013fb}
C.~Abs~da Cruz, P.~Chantrenne, R.~Gomes~de Aguiar~Veiga, M.~Perez, X.~Kleber.
\newblock Modified embedded-atom method interatomic potential and interfacial
  thermal conductance of Si-Cu systems: A molecular dynamics study.
 \newblock J. Appl. Phys. \textbf{113}, 023710 (2013)

\bibitem{Li:2019}
Z.~Li, S.~Xiong, C.~Sievers, Y.~Hu, Z.~Fan, N.~Wei, H.~Bao, S.~Chen,
  D.~Donadio, T.~Ala-Nissila, Influence of thermostatting on non-equilibrium
  molecular dynamics simulations of heat conduction in solids.
\newblock J. Chem. Phys. \textbf{151}, 234105 (2019)

\bibitem{Debye:1914}
P.~Debye, \emph{Zustandsgleichung und {Q}uantenhypothese mit einem {A}nhang
  \"{u}ber {W}\"{a}rmeleitung} (Teubner, Leipzig and Berlin, 1914), pp. 17--60.
\newblock In German

\bibitem{LennardJones:1931}
J.E. Lennard~Jones, Cohesion.
\newblock Proc. Phys. Soc. \textbf{43}, 461--482 (1931)

\bibitem{Plimpton:1995fc}
S.~Plimpton, {Fast Parallel Algorithms for Short-Range Molecular Dynamics}.
\newblock Journal of Computational Physics \textbf{117}(1), 1--19 (1995)

\bibitem{Parrinello:1980kx}
M.~Parrinello, A.~Rahman, {Crystal Structure and Pair Potentials: A
  Molecular-Dynamics Study}.
\newblock Physical Review Letters \textbf{45}(14), 1196--1199 (1980)

\bibitem{Parrinello:1981}
M.~Parrinello, A.~Rahman, J. Appl. Phys. \textbf{52}, 7182 (1981)

\bibitem{Peterson:1994cu}
R.B. Peterson, {Direct Simulation of Phonon- Mediated Heat Transfer in a Debye
  Crystal}.
\newblock Journal of Heat Transfer \textbf{116}, 815--822 (1994)

\bibitem{Lacroix:2005gm}
D.~Lacroix, K.~Joulain, D.~Lemonnier, {Monte Carlo transient phonon transport
  in silicon and germanium at nanoscales}.
\newblock Physical Review B \textbf{72}(6), 064305 (2005)

\bibitem{Peraud:2011ks}
J.P.M. P{\'e}raud, N.G. Hadjiconstantinou, {Efficient simulation of
  multidimensional phonon transport using energy-based variance-reduced Monte
  Carlo formulations}.
\newblock Physical Review B \textbf{84}, 205331 (2011)

\bibitem{Peraud:2012bj}
J.P.M. P{\'e}raud, N.G. Hadjiconstantinou, {An alternative approach to
  efficient simulation of micro/nanoscale phonon transport}.
\newblock Applied Physics Letters \textbf{101}, 153114 (2012)

\bibitem{Raleva:2017}
K.~Raleva, A.R. Shaik, D.~Vasileska, S.M.~Goodnick, \emph{Phonon Monte Carlo simulation}
  (Margan and Claypool Publishers, San Rafael CA, 2017), chap.~3

\bibitem{Gibson:2022}
G.W. Gibson, Monte {C}arlo simulated heat transport in semiconductor
  nanostructures.
\newblock Master's thesis, Laurentian University, Sudbury, ON (2022).
\newblock \url{https://zone.biblio.laurentian.ca/jspui/handle/10219/4005}

\bibitem{Jean:2014gw}
V.~Jean, S.~Fumeron, K.~Termentzidis, S.~Tutashkonko, D.~Lacroix, {Monte Carlo
  simulations of phonon transport in nanoporous silicon and germanium}.
\newblock Journal of Applied Physics \textbf{115}(2), 024304--13 (2014)

\bibitem{Esfarjani:2011}
K.~Esfarjani, G.~Chen, H.T. Stokes, Heat transport in silicon from
  first-principles calculations.
\newblock Phys. Rev. B \textbf{84}, 085204 (2011)

\bibitem{Regner:2013}
K.T. Regner, D.P. Sellan, Z.~Su, C.H. Amon, A.J.H. McGaughey, J.A. Malen,
  Broadband phonon mean free path contributions to thermal conductivity
  measured using frequency domain thermoreflectance.
\newblock Nat. Commun. \textbf{4}, 1640 (2013)

\bibitem{Casimir:1938}
H.B.G. Casimir, Note on the conduction of heat in crystals.
\newblock Physica \textbf{5}(6), 495--500 (1938)

\end{thebibliography}

\end{document}